\RequirePackage{ifpdf}
\ifpdf 
\documentclass[pdftex]{sigma}
\else
\documentclass{sigma}
\fi

\begin{document}

\allowdisplaybreaks

\renewcommand{\PaperNumber}{062}

\FirstPageHeading

\renewcommand{\thefootnote}{$\star$}

\ShortArticleName{Weakly Nonlocal Hamiltonian Structures: Lie Derivative and Compatibility}

\ArticleName{Weakly Nonlocal Hamiltonian Structures:\\
Lie Derivative and Compatibility\footnote{This paper is a
contribution to the Vadim Kuznetsov Memorial Issue `Integrable
Systems and Related Topics'. The full collection is available at
\href{http://www.emis.de/journals/SIGMA/kuznetsov.html}{http://www.emis.de/journals/SIGMA/kuznetsov.html}}}

\Author{Artur SERGYEYEV}

\AuthorNameForHeading{A. Sergyeyev}

\Address{Mathematical Institute, Silesian University in Opava,\\ Na
Ryb\-n\'\i{}\v cku 1, 746~01 Opava, Czech Republic}
\Email{\href{mailto:Artur.Sergyeyev@math.slu.cz}{Artur.Sergyeyev@math.slu.cz}}

\ArticleDates{Received December 15, 2006, in f\/inal form April
23, 2007; Published online April 26, 2007}

\Abstract{We show that under certain technical assumptions any weakly
nonlocal Hamiltonian structure compatible with a given nondegenerate
weakly nonlocal symplectic structure~$J$ can be written as the Lie
derivative of $J^{-1}$ along a suitably chosen nonlocal vector
f\/ield. Moreover, we present a new description for local Hamiltonian
structures of arbitrary order compatible with a given nondegenerate
local Hamiltonian structure of zero or f\/irst order, including
Hamiltonian operators of the Dubrovin--Novikov type.}


\Classification{37K10; 37K05}

\vspace{-2mm}

\section{Introduction}

Nonlinear integrable systems usually are bihamiltonian, i.e., possess
two compatible Hamiltonian structures. This ingenious discovery of Magri \cite{magri}
has naturally lead to an intense study of pairs
of compatible Hamiltonian structures both in f\/initely and inf\/initely many dimensions,
see e.g.\ \cite{bl,cooke,dor,li,mn, olv_eng2, serg1,vai} and references
therein.

Using the ideas from the Lichnerowicz--Poisson cohomology
theory \cite{li,vai} it can be shown \cite{dor,serg1} that under
certain minor technical assumptions all Hamiltonian structures
compatible with a given nondegenerate Hamiltonian
structure $P$ can be written as the Lie derivatives of~$P$ along
suitably chosen vector f\/ields. This allows for a
considerable reduction in the number of unknown functions: roughly
speaking, we deal with components of a vector f\/ield rather than
with those of a skew-symmetric tensor, and the number of the former is
typically much smaller than that of the latter, see e.g.\ \cite{serg1} 
for more details.
This idea works well for compatible pairs of f\/inite-dimensional
Hamiltonian structures \cite{serg1, smi} and of local Hamiltonian
operators of Dubrovin--Novikov type \cite{m1,serg1}, when the
corresponding vector f\/ields are local as well. \looseness=-1

In the present work we 
extend this approach to the
weakly nonlocal \cite{mn} Hamiltonian structures using
weakly nonlocal vector f\/ields. To this end we f\/irst generalize the
local homotopy formula (\ref{hom}) to weakly nonlocal symplectic
structures in Theorem~\ref{gam0} below. This enables us to
characterize large classes of Hamiltonian structures compatible with
a given weakly nonlocal symplectic structure using the weakly
nonlocal (co)vector f\/ields, i.e., elements of $\tilde{\mathcal{V}}$
(resp.\ $\tilde{\mathcal{V}}^*$), as presented below in
Theorems~\ref{hso} and \ref{hso2} and Corollaries~\ref{gam2},
\ref{gam3}, \ref{gam3loc}, and \ref{gam3a}.

The paper is organized as follows. In Section 2 we recall some basic
features of inf\/inite-dimensional Hamiltonian formalism. Section 3
contains the main theoretical results of the paper while Sections 4
and~5 deal with the particular cases of local Hamiltonian structures
of zero and f\/irst order where important simplif\/ications occur.
Finally, in Section 6 we brief\/ly discuss the results of the present work.

\section{Preliminaries}

Following \cite{dor, olv_eng2}, recall some basic aspects of
inf\/inite-dimensional Hamiltonian formalism
for the case of one independent variable $x\in B$ (usually $B=\mathbb{R}$
or $B=S^1$) and $n$ dependent variables.

We start with an algebra $\mathcal{A}_{j}$ of 
smooth functions of
$x,\boldsymbol{u},\boldsymbol{u}_{1},\dots,\allowbreak
\boldsymbol{u}_{j}$, where
$\boldsymbol{u}_{k}=(u_{k}^{1},\dots,u_{k}^{n})^{T}$ for $k>0$ are
$n$-component vectors from $\mathbb{R}^n$,
$\boldsymbol{u}_0\equiv\boldsymbol{u}\in M\subset\mathbb{R}^n$,
$M$ is an open domain in $\mathbb{R}^n$,
and the superscript $T$ indicates the transposed matrix. Set
$\mathcal{A}=\bigcup_{j=0}^{\infty}\mathcal{A}_{j}$. The elements of
$\mathcal{A}$ are called {\em local} functions.

Consider (see e.g.\ \cite{dor} and \cite{olv_eng2} and references
therein) a derivation of $\mathcal{A}$
\[
D\equiv D_x=\partial/\partial
x+\sum\limits_{j=0}^{\infty}\boldsymbol{u}_{j+1}\partial/\partial\boldsymbol{u}_j.
\]
and let $\mathrm{Im}\, D$ be the image of $D$ in $\mathcal{A}$, and
$\bar{\mathcal{A}}=\mathcal{A}/\mathrm{Im}\, D$. The space
$\bar{\mathcal{A}}$ is the counterpart the algebra of (smooth)
functions on a f\/inite-dimensional manifold in the standard de Rham
complex. Informally, $x$ can be thought of as a space variable and
$D$ as a total $x$-derivative, cf.\ e.g.\ \cite{olv_eng2}.
\looseness=-1

The canonical projection $\pi:\mathcal{A}\rightarrow
\bar{\mathcal{A}}$ is traditionally denoted by $\int dx$, and for
any $f,g\in\mathcal{A}$ we have
\[
\int f D(g) dx=-\int g D(f) dx.
\]
The quantity $\mathcal{F}=\int f dx$ should not be confused
with a nonlocal variable $D^{-1}(f)$: these are dif\/ferent objects.
Informally, $\int f dx$ can be thought of as $\int_{B} f dx$,
i.e., this is, roughly speaking, a~def\/inite $x$-integral, and
$D^{-1}(f)$ is a formal indef\/inite $x$-integral. If $f\not\in
\mathrm{Im} D $ then $D^{-1}(f)\not\in\mathcal{A}$, and we need to
augment $\mathcal{A}$ to include a nonlocal variable $\omega$ such
that $D(\omega)=f$ and to extend the action of $D$ accordingly, see
below for further details.

The generalized Leibniz rule \cite{s, mik1, mik, olv_eng2}
\begin{equation}\label{lei} a D^{i}\circ b D^{j} =a
\sum\limits_{q=0}^{\infty}{\displaystyle \frac{i(i-1)\cdots
(i-q+1)}{q!}}D^{q}(b)D^{i+j-q} \end{equation} turns the space
$\mathrm{Mat}_{q}(\mathcal{A})[\![D^{-1}]\!]$ of formal series in
powers of $D$ of the form $L=\sum
_{j=-\infty}^{k}h_{j}D^{j}$, where $h_{j}$ are $q\times q$ matrices
with entries from $\mathcal{A}$, into an algebra, and the commutator
$\left[ P, Q \right]=P \circ Q
- Q \circ P$ further makes
$\mathrm{Mat}_{q}(\mathcal{A})[\![D^{-1}]\!]$ into a Lie algebra. In
what follows we shall often omit the composition sign $\circ$ (for
instance, we shall write  $K L$ instead of $K\circ L$) wherever this
does not lead to a possible confusion.

The {\em degree} $\deg L$ of formal series $L=\sum
_{j=-\infty}^{k}h_{j}D^{j}
\in\mathrm{Mat}_{q}(\mathcal{A})[\![D^{-1}]\!]$
is \cite{s, mik1, mik, olv_eng2} the grea\-test integer $m$ such that $h_{m}\neq 0$. If,
moreover, $\det h_m\neq 0$ we shall call $L$ {\em nondegenerate},
and then there exists a unique formal series $L^{-1}\in
\mathrm{Mat}_{q}(\mathcal{A})[\![D^{-1}]\!]$ such that $L^{-1} \circ
L=L\circ L^{-1} =\mathbb{I}_q$, where $\mathbb{I}_q$ stands for the
$q\times q$ unit matrix. For any
$L=\sum_{j=-\infty}^{m}h_{j}D^{j}\in\mathrm{Mat}_{q}(\mathcal{A})[\![D^{-1}]\!]$
let $L_+=\sum_{j=0}^{m}h_{j}D^{j}$ denote its dif\/ferential part,
$L_-=\sum_{j=-\infty}^{-1}h_{j}D^{j}$ its nonlocal part (so
$L_-+L_+=L$),
and let $L^{\dagger}=\sum
_{j=-\infty}^{m}(-D)^{j}\circ h_{j}^{T}$ stand for the formal
adjoint of $L$, see e.g.\ \cite{s, mik1, mik, olv_eng2}.
A formal series $L$ is said to be {\em skew-symmetric} if
$L^{\dagger}=-L$. As usual, an $L\in
\mathrm{Mat}_{q}(\mathcal{A})[\![D^{-1}]\!]$ is said to be a {\em
purely differential} (or just {\em differential}) operator if
$L_-=0$. \looseness=-1

Let $\mathcal{A}^q$ be the space of $q$-component functions with
entries from $\mathcal{A}$, no matter whether they are interpreted
as column or row vectors. For any $\vec f\in\mathcal{A}^{q}$
def\/ine (see e.g.\ \cite{ff}) its directional derivative as
\[
\vec f\,{}'=\sum\limits_{i=0}^{\infty}\partial \vec
f/\partial\boldsymbol{u}_{i}D^{i}.
\]
We shall also need the operator of variational derivative (see e.g.\
\cite{bl, dor, olv_eng2,v2})
\[
\delta/\delta\boldsymbol{u}=\sum\limits_{j=0}^{\infty}(-D)^{j}\circ
\partial/\partial\boldsymbol{u}_j.
\]

Following \cite{mn}, an
$L\in\mathrm{Mat}_{q}(\mathcal{A})[\![D^{-1}]\!]$ is called {\em
weakly nonlocal} if there exist $\vec f_\alpha\in\mathcal{A}^q$,
$\vec g_\alpha\in\mathcal{A}^q$ and $k\in\mathbb{N}$ such that
$L_{-}=\sum_{\alpha=1}^{k} \vec f_\alpha\otimes D^{-1}\circ \vec
g_\alpha$.
Nearly all known today 
Hamiltonian and symplectic operators in (1+1) dimensions
are weakly nonlocal, cf.\ e.g.\ \cite{wang}. \looseness=-1
Recall
that an operator of the form $L=\vec f\otimes D^{-1}\circ \vec g$
acts on an $\vec h \in\mathcal{A}^q$ as follows:
\[
L(\vec h)=\big(D^{-1}\big(\vec g \cdot \vec h\big)\big) \vec
f,
\]
where ``$\cdot$''  denotes the standard Euclidean scalar product in
$\mathcal{A}^q$.

Denote by $\mathcal{V}$ the space of $n$-component columns with
entries from $\mathcal{A}$.
The commutator
$[\boldsymbol{P},\boldsymbol{Q}]=\boldsymbol{Q}'[\boldsymbol{P}]-\boldsymbol{P}'[\boldsymbol{Q}]$
turns $\mathcal{V}$ into a Lie algebra, see e.g.\ \cite{
bl, ff, mik1, olv_eng2}.  The
Lie derivative of $\boldsymbol{R}\in\mathcal{V}$ along
$\boldsymbol{Q}\in\mathcal{V}$ reads
$L_{\boldsymbol{Q}}(\boldsymbol{R})=[\boldsymbol{Q},\boldsymbol{R}]$, see
e.g.\ \cite{bl, dor, wangth, olv_eng2}.
The natural dual of $\mathcal{V}$ is the space $\mathcal{V}^*$ of
$n$-component rows with entries from $\mathcal{A}$.

The canonical pairing of $\mathcal{V}$ and $\mathcal{V}^*$ is given
by the formula (see e.g.\ \cite{dor,wang})
\begin{equation}\label{pairg}
\langle\boldsymbol{\gamma},\boldsymbol{Q}\rangle=\int
(\boldsymbol{\gamma}\cdot\boldsymbol{Q})dx, \end{equation} where
$\boldsymbol{\gamma}\in\mathcal{V}^*,\boldsymbol{Q}\in\mathcal{V}$,
and ``$\cdot$" here and below refers to the standard Euclidean scalar product of
the $n$-component vectors.

For $\boldsymbol{\gamma}\in\mathcal{V}^*$ def\/ine \cite{bl, dor,
wangth} its Lie derivative along $\boldsymbol{Q}\in\mathcal{V}$ as
$L_{\boldsymbol{Q}}(\boldsymbol{\gamma})
=\boldsymbol{\gamma}'[\boldsymbol{Q}]-(\boldsymbol{Q}')^{\dagger}(\boldsymbol{\gamma})$,
see e.g.\ \cite{dor, wangth} for further details.

For $\boldsymbol{Q}\in\mathcal{V}$ and
$L=\sum_{j=-\infty}^{m}h_{j}D^{j}$ we set
$L'[\boldsymbol{Q}]=\sum_{j=-\infty}^{m}h'_{j}[\boldsymbol{Q}]D^{j}$.

If $\boldsymbol{Q}\in\mathcal{V}$ and $\boldsymbol{\gamma}\in\mathcal{V}^*$ then 
we have \cite{olv_eng2}
$\delta(\boldsymbol{Q}\cdot\boldsymbol{\gamma})/\delta\boldsymbol{u}=(\boldsymbol{Q}')^\dagger(\boldsymbol{\gamma})
+(\boldsymbol{\gamma}')^\dagger(\boldsymbol{Q})$. Hence if
$(\boldsymbol{\gamma}')^\dagger(\boldsymbol{Q})-\boldsymbol{\gamma}'[\boldsymbol{Q}]=0$
then we obtain \cite{serg2}
\begin{equation}\label{lievar}
L_{\boldsymbol{Q}}(\boldsymbol{\gamma})=\delta(\boldsymbol{Q}\cdot\boldsymbol{\gamma})/\delta\boldsymbol{u}.
\end{equation}

For weakly nonlocal $R:\mathcal{V}\rightarrow\mathcal{V}$,
$J:\mathcal{V}\rightarrow\mathcal{V}^*$,
$P:\mathcal{V}^*\rightarrow\mathcal{V}$,
$N:\mathcal{V}^*\rightarrow\mathcal{V}^*$
def\/ine \cite{ff} their Lie derivatives along a
$\boldsymbol{Q}\in\mathcal{V}$ as follows:
$L_{\boldsymbol{Q}}(R)=R'[\boldsymbol{Q}]-[\boldsymbol{Q}',R]$,
$L_{\boldsymbol{Q}}(N)=N'[\boldsymbol{Q}]+[\boldsymbol{Q}'^\dagger,N]$,
$L_{\boldsymbol{Q}}(P)=P'[\boldsymbol{Q}] -P\circ
\boldsymbol{Q}'-\boldsymbol{Q}'^\dagger \circ P$,
$L_{\boldsymbol{Q}}(J)=J'[\boldsymbol{Q}] +J\circ
\boldsymbol{Q}'+\boldsymbol{Q}'^\dagger \circ J$. Here and below we
do {\em not} assume $R$ and $J$ to be def\/ined on the whole of
$\mathcal{V}$, respectively $P$ and $N$
on the whole of $\mathcal{V}^*$.

We shall call an operator $J:\mathcal{V}\rightarrow\mathcal{V}^*$ (respectively
$P:\mathcal{V}^*\rightarrow\mathcal{V}$) {\em formally skew-symmetric} if it is
skew-symmetric when considered as a formal series, i.e., $J^\dagger=-J$ (respectively
$P^\dagger=-P$).

Recall that the proper way
to extend the concept of
the f\/inite-dimensional
Hamiltonian structure to evolutionary systems of PDEs
in (1+1) dimensions is the following one.
A formally skew-symmetric operator $P:
\mathcal{V}^*\rightarrow\mathcal{V}$ is {\em Hamiltonian} \cite{dor}
(or {\em implectic} \cite{ff}) if its Schouten bracket with itself vanishes:
$[P,P]=0$. The Schouten bracket $[\cdot,\cdot]$ is given by the formula 
\looseness=-1
\begin{equation}\label{sch}
[H,K](\boldsymbol\chi_1,\boldsymbol\chi_2,\boldsymbol\chi_3)=
\langle H L_{K
\boldsymbol\chi_1}(\boldsymbol\chi_2),\boldsymbol\chi_3\rangle
+\langle K L_{H
\boldsymbol\chi_1}(\boldsymbol\chi_2),\boldsymbol\chi_3 \rangle +
\mbox{cycle}(1,2,3), \end{equation} where
$\boldsymbol\chi_i\in\mathcal{V}^*$ and $\langle,\rangle$ is given
by (\ref{pairg}), see e.g.\ \cite{dor}.
Throughout the rest of the paper $[\cdot,\cdot]$ will denote
the Schouten bracket rather than the commutator.
\looseness=-1

Two Hamiltonian operators are said to be {\em compatible} \cite{ff}
(or to form a {\em Hamiltonian pair} \cite{dor}) if any linear
combination thereof is again a Hamiltonian operator. Note that the
Hamiltonian operators are compatible if and only if their Schouten
bracket vanishes \cite{dor}.

The Poisson bracket $\{,\}_P$ associated with a Hamiltonian operator
$P$ is (see e.g.\ \cite{dor, olv_eng2})
a~mapping from $\bar{\mathcal{A}} \times\bar{\mathcal{A}}$ to
$\bar{\mathcal{A}}$ given by the formula{\samepage \begin{equation}\label{bra}
\{\mathcal{F},\mathcal{G}\}_P=\int dx \delta\mathcal{F}
P(\delta\mathcal{G}) \end{equation} for any
$\mathcal{F},\mathcal{G}\in\bar{\mathcal{A}}$. Here we set
$\delta\mathcal{F}\stackrel{\mathrm{def}}{=}\delta
f/\delta\boldsymbol{u}$ for any $\mathcal{F}=\int
f dx\in\bar{\mathcal{A}}$.}

A formally skew-symmetric operator $J:\mathcal{V}\rightarrow\mathcal{V}^*$ is
{\em symplectic} \cite{ff} if
\begin{equation}\label{symp} \langle
J'[\boldsymbol{P}]\boldsymbol{Q},\boldsymbol{R}\rangle +\langle
J'[\boldsymbol{Q}]\boldsymbol{R},\boldsymbol{P}\rangle +\langle
J'[\boldsymbol{R}]\boldsymbol{P},\boldsymbol{Q}\rangle=0
\end{equation} for any
$\boldsymbol{P},\boldsymbol{Q},\boldsymbol{R}\in\mathcal{V}$.

Following the tradition established in the literature
we shall sometimes speak of Hamiltonian (or symplectic) structures
rather than of Hamiltonian (or symplectic) operators, even though the
latter terms are equivalent with the former.

We shall call a Hamiltonian or symplectic operator {\em nondegenerate} if
it is nondegenerate as a~formal series in powers of $D$. A
nondegenerate operator $P:\mathcal{V}^*\rightarrow\mathcal{V}$ is
Hamiltonian if and only if $P^{-1}$ is symplectic. Following \cite{ff}, and
in contrast with a number of other references,
in what follows we do {\em not} assume
symplectic operators to be {\em a priori} nondegenerate. 
\looseness=-1

We have the following 
homotopy formula (see \cite[Ch.~5]{olv_eng2}
and \cite{dor, oevth} for details): if
$J:\mathcal{V}\rightarrow\mathcal{V}^*$ is a {\em differential}
symplectic operator and $M \times B$ is a star-shaped domain (recall
that $M$ and $B$ are domains of values of $\boldsymbol{u}$ and $x$,
respectively) then we have
$J=\boldsymbol{\zeta}'-\boldsymbol{\zeta}'^\dagger$ for
\begin{equation}\label{hom} \boldsymbol{\zeta}=\int_0^1
(J(\boldsymbol{u}))[\lambda\boldsymbol{u}] {d\lambda}.
\end{equation}
Here $J(\boldsymbol{u})$ means the result of action of the
dif\/ferential operator $J$ on the vector $\boldsymbol{u}$, and for
any $f\in\mathcal{A}$ the quantity $f[\lambda\boldsymbol{u}]$ is
def\/ined as follows: if
$f=f(x,\boldsymbol{u},\dots,\boldsymbol{u}_k)$ then
\[
f[\lambda\boldsymbol{u}]\stackrel{\mathrm{def}}{=}f(x,\lambda\boldsymbol{u},\dots,\lambda\boldsymbol{u}_k).
\]
In what follows we make the {\em blanket assumption} that $M \times
B$ is a star-shaped domain so that (\ref{hom}) is automatically
valid.

In order to see how (\ref{hom}) works, consider the following simple example.
Let $J=D$. Then we have $J(\boldsymbol{u})=D(\boldsymbol{u})=\boldsymbol{u}_1$,
and therefore $(J(\boldsymbol{u}))[\lambda\boldsymbol{u}]=\lambda\boldsymbol{u}_1$.
By (\ref{hom}) we obtain $\boldsymbol{\zeta}=\boldsymbol{u}_1/2$ and indeed
the equality $J=\boldsymbol{\zeta}'-\boldsymbol{\zeta}'^\dagger$ holds, as desired.

Note that the proper geometrical framework for the above results is
provided by the 
formal calculus of variations, and we
refer the interested reader to \cite{v2, dor, olv_eng2, wangth} and
references therein for further details.

Our immediate goal is to
generalize (\ref{hom}) to the case when the matrix
operator $J$
is weakly nonlocal rather than purely dif\/ferential, see Theorem~\ref{gam0}
below. However, we shall need
a few more def\/initions and known results in order to proceed.

A symplectic operator $J$ is {\em compatible} \cite{ff} with a
Hamiltonian operator $\tilde P$ if $J \tilde P J$ is again
symplectic. If the symplectic operator $J$ is an inverse of a
Hamiltonian operator $P$, then the compatibility of $J$ and
$\tilde P$ is equivalent to that of $P$ and $\tilde P$. In fact, a
more general assertion holds.
\looseness=-1
\begin{lemma}\label{lemcomp}
Consider a nondegenerate Hamiltonian operator $P$ and a formally
skew-symmetric operator $\tilde P:\mathcal{V}^*\rightarrow \mathcal{V}$
which is not necessarily Hamiltonian.
Their Schouten bracket
vanishes ($[P, \tilde P]=0$) if and only if the operator 
$P^{-1} \tilde P P^{-1}$ is symplectic.
\end{lemma}
\begin{proof}[Sketch of proof]
By (\ref{symp}), 
the operator $\widetilde J=P^{-1}\tilde P P^{-1}$
is symplectic if and only if \begin{equation}\label{symp1} \langle
\widetilde
J'[\boldsymbol{X}_1]\boldsymbol{X}_2,\boldsymbol{X}_3\rangle
+\langle \widetilde
J'[\boldsymbol{X}_2]\boldsymbol{X}_3,\boldsymbol{X}_1\rangle
+\langle \widetilde
J'[\boldsymbol{X}_3]\boldsymbol{X}_1,\boldsymbol{X}_2\rangle=0.
\end{equation}
Let $\boldsymbol{X}_i=P\boldsymbol{\chi}_i$,
$\boldsymbol\chi_i\in\mathcal{V}$. By equation~(4.12) and Proposition 4.3
of \cite{vai} which are readily seen to be applicable in the
inf\/inite-dimensional case as well, we have
\[
[P,\tilde P](\boldsymbol\chi_1,\boldsymbol\chi_2,\boldsymbol\chi_3)
=\langle \widetilde J'[\boldsymbol{X}_1]
\boldsymbol{X}_2,\boldsymbol{X}_3\rangle +\langle \widetilde
J'[\boldsymbol{X}_2]\boldsymbol{X}_3,\boldsymbol{X}_1\rangle
+\langle \widetilde
J'[\boldsymbol{X}_3]\boldsymbol{X}_1,\boldsymbol{X}_2\rangle,
\]
and the result follows.
\end{proof}

Note also the following 
easy corollary of Theorem 1 of \cite{mal}.
\begin{theorem}\label{fso}
Let $\varepsilon_\alpha$ be arbitrary nonzero constants, and
$\psi_\alpha\in\mathcal{A}$ be local functions such that
$\delta\psi_\alpha/\delta\boldsymbol{u}\neq 0$ for all
$\alpha=1,\dots,q$. Then the operator \begin{equation}\label{so1}
J=\sum\limits_{\alpha=1}^q \varepsilon_\alpha
\displaystyle\frac{\delta\psi_\alpha}{\delta\boldsymbol{u}}\otimes
D^{-1}\circ \frac{\delta\psi_\alpha}{\delta\boldsymbol{u}}
\end{equation} is symplectic.
\end{theorem}

We now need to extend $\mathcal{A}$, $\mathcal{V}$ and
$\mathcal{V}^*$ to include {\em weakly nonlocal elements}. First of
all, a~$q$-com\-ponent vector function $\vec f$ is said to be {\em
weakly nonlocal} if there exist a nonnegative integer $s$ and $\vec
f_0\in\mathcal{A}^q$, $\vec f_\alpha\in\mathcal{A}^q$,
$K_\alpha\in\mathcal{A}$, $\alpha=1,\dots,s$ such that $\vec f$ can
be written as \begin{equation}\label{wnlf} \vec f=\vec
f_0+\sum\limits_{\alpha=1}^s \vec f_\alpha D^{-1}(K_\alpha),
\end{equation} where $\vec f_\alpha$ are {\em linearly independent}
over $\mathcal{A}$ for $\alpha=1,\dots,s$,
$\delta K_\alpha/\delta \boldsymbol{u}\neq 0$,
$\alpha=1,\dots,s$, and $K_\alpha$ are linearly independent over the 
constants.

We shall denote the space of weakly nonlocal $q$-component vectors in
the sense of above def\/inition by $\tilde{\mathcal{A}}^q$;
$\tilde{\mathcal{V}}$ (resp.\ $\tilde{\mathcal{V}}^*$) will stand
for the space of $n$-component columns (resp.\ rows) with entries
from $\tilde{\mathcal{A}}\equiv \tilde{\mathcal{A}}^1$.
The def\/inition of directional derivative is 
extended
to $\tilde{\mathcal{A}}^q$ as follows: for $\vec f$ of the form
(\ref{wnlf}) we set
\[
\vec f'=\vec f_0\,'+\sum\limits_{\alpha=1}^s \left( D^{-1}(K_\alpha)
\vec f_\alpha\,' +\vec f_\alpha D^{-1}\circ K'_\alpha\right).
\]
Moreover, the def\/initions of directional derivative and the Lie
derivative along $\boldsymbol{Q}\in\mathcal{V}$ readily extend to
the elements of $\tilde{\mathcal{V}}$. In the present paper we adopt
a relatively informal approach to nonlocal variables in spirit of
\cite{ff3}. For a more rigorous approach to nonlocal
symmetries see e.g. \cite{v2, as-sb} and references therein.

We shall call a weakly nonlocal Hamiltonian operator $P$ {\em
normal} if for any $\boldsymbol{Q}\in\tilde{\mathcal{V}}$ the
condition $L_{\boldsymbol{Q}}(P)=0$ implies that
$\boldsymbol{Q}\in\mathcal{V}$.

\section{Main results}
We start with the following nonlocal generalization of
the homotopy formula (\ref{hom}).

\begin{theorem}\label{gam0}
Let $J: \mathcal{V}\rightarrow\mathcal{V}^*$ be a weakly nonlocal
formally skew-symmetric 
operator. Suppose that
there exist $\varepsilon_\alpha$ and local $H_\alpha$ such that
$\varepsilon_\alpha^2=1$ (i.e., $\varepsilon_\alpha=\pm 1$) and
we have
\[
J_{-}=\sum\limits_{\alpha=1}^q \varepsilon_\alpha \delta
H_\alpha/\delta\boldsymbol{u} \otimes D^{-1}\circ\delta
H_\alpha/\delta\boldsymbol{u}.
\]

Then the operator $J$ is symplectic if and only if there exists a
local $\boldsymbol{\gamma}_0\in\mathcal{V}^*$ such that
we have $J=\boldsymbol{\gamma}\,'-(\boldsymbol{\gamma}\,')^\dagger$
for
\begin{equation}\label{gameq0}
\boldsymbol{\gamma}=\boldsymbol{\gamma}_0 +\displaystyle\frac12
\sum\limits_{\alpha=1}^q \varepsilon_\alpha \delta
H_\alpha/\delta\boldsymbol{u} D^{-1}(H_\alpha).
\end{equation}
\end{theorem}
\begin{proof} If there exists $\boldsymbol{\gamma}_0$ such that $\boldsymbol{\gamma}$ (\ref{gameq0})
satisf\/ies 
\begin{equation}\label{jdg}
J=\boldsymbol{\gamma}\,'-(\boldsymbol{\gamma}\,')^\dagger
\end{equation}
then $J$ is obviously symplectic.

Now assume that
$J$ is symplectic and construct a suitable $\boldsymbol{\gamma}_0$ such that
$\boldsymbol{\gamma}$ (\ref{gameq0}) satisf\/ies (\ref{jdg}).
Let
$\widetilde{\boldsymbol{\gamma}}=\boldsymbol{\gamma}-\boldsymbol{\gamma}_0$.
We readily see that we have \looseness=-1
\begin{equation}\label{jg}
\big(\widetilde{\boldsymbol{\gamma}}\,'
-\widetilde{\boldsymbol{\gamma}}\,'^\dagger\big)_-=J_-.
\end{equation}
On the other hand,
$\big(\widetilde{\boldsymbol{\gamma}}'
-\widetilde{\boldsymbol{\gamma}}'^\dagger\big)$ obviously is a
symplectic operator and therefore so is
\[\tilde J=J-\big(\widetilde{\boldsymbol{\gamma}}\,'-\widetilde{\boldsymbol{\gamma}}\,'^\dagger\big).
\]
By virtue of (\ref{jg}) we have $\tilde J_-=0$, i.e., $\tilde J$ is purely
dif\/ferential.
Let
\[\boldsymbol{\gamma}_0=\int_0^1
(\tilde J(\boldsymbol{u}))[\lambda\boldsymbol{u}] {d\lambda}.\]

Clearly, this $\boldsymbol{\gamma}_0$ is
local \cite{dor}, 
and by (\ref{hom}) we have $\tilde
J=\boldsymbol{\gamma}\,'_0-(\boldsymbol{\gamma}\,'_0)^\dagger$.
Hence $\boldsymbol{\gamma}$ (\ref{gameq0}) satisf\/ies (\ref{jdg}),
and the result follows.
\end{proof}

Theorem~\ref{gam0} means that the existence of a (not necessarily
globally def\/ined) weakly nonlocal~$\boldsymbol{\gamma}$ such that (\ref{jdg})
holds
is a necessary and suf\/f\/icient condition for a weakly nonlocal $J$ to
be symplectic. An important feature of this result is that the
nonlocal terms in $\boldsymbol{\gamma}$ are uniquely determined by the
structure of nonlocal terms in $J$, so in fact we only need to determine a
{\em local} $\boldsymbol{\gamma}_0$.
\looseness=-1

Combining Lemma~\ref{lemcomp} and Theorem~\ref{gam0} we arrive at the
following results.
\begin{corollary}\label{gam}
Let $P$ be a nondegenerate Hamiltonian operator and
$\tilde P:\mathcal{V}^*\rightarrow\mathcal{V}$ be a formally
skew-symmetric operator such that $P^{-1}\tilde P P^{-1}$ is weakly
nonlocal and there exist $\varepsilon_\alpha=\pm 1$ and local
$F_\alpha$ such that \begin{equation}\label{wn} P^{-1}\tilde P
P^{-1}=\sum\limits_{\alpha=1}^s \varepsilon_\alpha \delta
F_\alpha/\delta\boldsymbol{u} \otimes D^{-1}\circ\delta
F_\alpha/\delta\boldsymbol{u}. \end{equation}

Then $[P,\tilde P]=0$ if and only if there exists
a local $\boldsymbol{\gamma}_0\in\mathcal{V}^*$
such that
\begin{equation}\label{gameq} \boldsymbol{\gamma}=\boldsymbol{\gamma}_0
+\displaystyle\frac12 \sum\limits_{\alpha=1}^s \varepsilon_\alpha
\delta F_\alpha/\delta\boldsymbol{u} D^{-1}(F_\alpha)\end{equation}
satisfies $P^{-1}\tilde P
P^{-1}=\boldsymbol{\gamma}\,'-(\boldsymbol{\gamma}\,')^\dagger$.
\end{corollary}

\begin{corollary}\label{gam2} Under
the assumptions of Corollary~{\rm \ref{gam}} suppose that $P$
is a normal 
weakly nonlocal Hamiltonian operator of the form
\begin{equation}\label{p} P=\sum\limits_{m=0}^{\bar
p}a_{m}D^{m}+\sum \limits_{\rho=1}^{\bar
q}\bar\epsilon_\rho\boldsymbol{G}_{\rho}\otimes D^{-1}\circ
\boldsymbol{G}_{\rho},
\end{equation} where $a_m$ are $n\times n$  matrices with entries from
$\mathcal{A}$, $\bar\epsilon_\rho$ are arbitrary nonzero
constants, $\boldsymbol{G}_{\rho}\in\mathcal{V}$, and we have
\begin{equation}\label{loctau0}
L_{\boldsymbol{G}_\rho}(\delta F_{\alpha}/\delta\boldsymbol{u})=0,
\qquad \alpha=1,\dots,s,\quad \rho=1,\dots,\bar p. \end{equation}

Then $[P,\tilde P]=0$ if and only if there exists a weakly nonlocal
$\boldsymbol{\tau}\in\tilde{\mathcal{V}}$ such that
$\tilde P=L_{\boldsymbol{\tau}}(P)$.
\end{corollary}

\begin{proof} 
Under the assumptions of Corollary~\ref{gam} let
$\boldsymbol{\tau}=-P\boldsymbol{\gamma}+\boldsymbol{Q}$, where
$\boldsymbol{\gamma}$ is given by (\ref{gameq}) and~$\boldsymbol{Q}$
satisf\/ies $L_{\boldsymbol{Q}}(P)=0$.
Then we have $\tilde P=L_{\boldsymbol{\tau}}(P)$,
cf.\ proof of Proposition 3 in \cite{serg1}.

The Hamiltonian operator $P$ is
normal by assumption, and hence $\boldsymbol{Q}$ is
local, i.e., $\boldsymbol{Q}\in\mathcal{V}$.
Hence the only nonlocal terms in $\boldsymbol{\tau}$
originate from $-P\boldsymbol{\gamma}$ and read
\[
-\displaystyle\frac12 \sum\limits_{\alpha=1}^s \varepsilon_\alpha
D^{-1}(F_\alpha) P(\delta F_\alpha/\delta\boldsymbol{u})+ \frac12
\sum\limits_{\alpha=1}^q \sum\limits_{\rho=1}^{\bar q}
\varepsilon_\alpha\bar \varepsilon_\rho \boldsymbol{G}_\rho D^{-1}\left(\left(D^{-1}\left(\delta
F_\alpha/\delta\boldsymbol{u}\cdot \boldsymbol{G}_\rho\right)\right)
F_\alpha\right).
\]

Now, the expressions $D^{-1}(\left(\delta
F_\alpha/\delta\boldsymbol{u}\cdot \boldsymbol{G}_\rho\right))$ are
in fact local. Indeed, by (\ref{lievar}) the conditions~(\ref{loctau0}) are equivalent to
\begin{equation}\label{aaa} \delta(\boldsymbol{G}_\rho\cdot\delta
F_{\alpha}/\delta\boldsymbol{u})/\delta\boldsymbol{u}=0, \qquad
\alpha=1,\dots,q,\quad \rho=1,\dots,\bar q.
\end{equation}
In turn, (\ref{aaa}) implies that $(\boldsymbol{G}_\rho\cdot\delta
F_{\alpha}/\delta\boldsymbol{u})\in\mathrm{Im} D$, as desired.

Hence $P(\delta F_\alpha/\delta\boldsymbol{u})$ and
$\left(D^{-1}\left(\delta F_\alpha/\delta\boldsymbol{u}\cdot
\boldsymbol{G}_\rho\right)\right) F_\alpha$ are local, and
$\boldsymbol{\tau}$ is weakly nonlocal.

On the other hand, if there exists a weakly nonlocal
$\boldsymbol{\tau}$ such that $\tilde P=L_{\boldsymbol{\tau}}(P)$ then we have
$[P,\tilde P]=0$, cf.\ the proof of Proposition 7.8 of \cite{dor} or
equation (4) of \cite{serg1},
and the result follows.
\end{proof}

The above two results are more than a mere test of whether a given
$\tilde P$ has a zero Schouten bracket with $P$ (and, in particular,
whether the Hamiltonian operators $P$ and $\tilde P$ are
compa\-tible). In particular, Corollary~\ref{gam2} shows that if $P$
is purely dif\/ferential and normal then, under certain technical
assumptions that appear to hold in all interesting examples, all
weakly nonlocal Hamiltonian operators compatible with $P$ can be
written in the form $L_{\boldsymbol{\tau}}(P)$ for suitably chosen
weakly nonlocal $\boldsymbol{\tau}$. \looseness=-1

Therefore, we can {\em search} for Hamiltonian operators compatible
with $P$ by picking a gene\-ral weakly nonlocal $\boldsymbol{\tau}$
and requiring
the operator $L_{\boldsymbol{\tau}}(P)$ to be Hamiltonian.
Clearly, 
we have considerably fewer unknown functions to determine than if we
would just assume that $\tilde P$ is weakly nonlocal and
formally skew-symmetric and then require $\tilde P$ to be a Hamiltonian
operator compatible with $P$.
\looseness=-1

It is natural to ask under which conditions the operator
$P^{-1}\tilde P P^{-1}$ meets the requirements of
Corollary~\ref{gam}.
To this end consider f\/irst a weakly nonlocal operator of the form
\begin{equation}\label{so} J=\sum\limits_{m=1}^p b_m D^m+
\sum\limits_{\alpha=1}^q \varepsilon_\alpha
\displaystyle\frac{\delta\psi_\alpha}{\delta\boldsymbol{u}}\otimes
D^{-1}\circ \frac{\delta\psi_\alpha}{\delta\boldsymbol{u}},
\end{equation} where $b_m$ are $n\times n$  matrices with entries
from $\mathcal{A}$, $\varepsilon_\alpha$ are arbitrary nonzero
constants, and $\psi_\alpha\in\mathcal{A}$ are local functions.

In what follows we assume without loss of generality that
$\delta\psi_\alpha/\delta\boldsymbol{u}$, $\alpha=1,\dots,q$, are linearly independent
over the 
constants. We have the following well-known result.

\begin{lemma}\label{cas}Let $J: \mathcal{V}\rightarrow\mathcal{V}^*$ be a nondegenerate operator of the form \eqref{so}.
If $P=J^{-1}$ is a~purely differential operator then we have
\begin{equation}\label{casi}
P\left(\displaystyle\frac{\delta\psi_\alpha}{\delta\boldsymbol{u}}\right)=0,\qquad
\alpha=1,\dots,q. \end{equation} In particular, if $J$ is symplectic
then $\int \psi_\alpha dx$ are Casimir functionals for the bracket
$\{,\}_P$.
\end{lemma}
\begin{proof} We have
\[
J(0)=\sum\limits_{\alpha=1}^q c_\alpha \varepsilon_\alpha
\displaystyle\frac{\delta\psi_\alpha}{\delta\boldsymbol{u}},
\]
where $c_\alpha$ are arbitrary constants. Acting by $P=J^{-1}$ on
the left- and right-hand side of this equation yields
\[
\sum\limits_{\alpha=1}^q c_\alpha \varepsilon_\alpha
P\left(\displaystyle\frac{\delta\psi_\alpha}{\delta\boldsymbol{u}}\right)=0,
\]
and since $c_\alpha$ are arbitrary we obtain (\ref{casi}).
\end{proof}

Further let $\tilde P$ be a weakly nonlocal 
formally skew-symmetric operator 
of the form \begin{equation}\label{tp} \tilde
P=\sum\limits_{m=0}^{\tilde p}\tilde{a}_{m}D^{m}+\sum
\limits_{\rho=1}^{\tilde
q}\tilde\varepsilon_\rho\boldsymbol{Y}_{\rho}\otimes D^{-1}\circ
\boldsymbol{Y}_{\rho}, \end{equation} where $\tilde a_m$ are
$n\times n$ matrices with entries from $\mathcal{A}$ and
$\tilde\varepsilon_\rho$ are arbitrary nonzero constants.

\begin{theorem}\label{hso}Let $J$ be a weakly nonlocal
symplectic operator of the form \eqref{so}
and $\tilde P:  \mathcal{V}^*\rightarrow\mathcal{V}$
be a weakly nonlocal 
formally skew-symmetric operator 
of the form \eqref{tp}.
Suppose that there exist local functions $H_\rho$ and $K_\alpha$
such that 
\begin{gather}\label{loca0} J \boldsymbol{Y}_\rho=\delta
H_\rho/\delta\boldsymbol{u},\qquad\rho=1,\dots,\tilde q,
\qquad\mbox{and}\\ J \tilde
P(\delta\psi_\alpha/\delta\boldsymbol{u})=\delta
K_\alpha/\delta\boldsymbol{u},
\qquad \alpha=1,\dots,q,\nonumber
 \end{gather} Then $J\tilde P J$ is weakly nonlocal and we have
\begin{gather} (J\tilde P J)_-=\sum\limits_{\alpha=1}^q
\varepsilon_\alpha \left(\displaystyle\frac{\delta
K_\alpha}{\delta\boldsymbol{u}}\otimes D^{-1}\circ \frac{\delta
\psi_\alpha}{\delta\boldsymbol{u}}+\frac{\delta
\psi_\alpha}{\delta\boldsymbol{u}}\otimes D^{-1}\circ \frac{\delta
K_\alpha}{\delta\boldsymbol{u}}\right)\nonumber\\
\phantom{(J\tilde P J)_-=}{} - \sum
\limits_{\rho=1}^{\tilde q}\tilde\epsilon_\rho \frac{\delta
H_\rho}{\delta\boldsymbol{u}}\otimes D^{-1}\circ \frac{\delta
H_\rho}{\delta\boldsymbol{u}}. \label{jpj}
\end{gather}
Moreover, 
the operator $J\tilde P J$ is symplectic if and only if there exists
a local $\boldsymbol{\gamma}_0\in\mathcal{V}^*$ such that
\begin{equation}\label{gameq2}
\boldsymbol{\gamma}=\boldsymbol{\gamma}_0 -\displaystyle\frac12 \sum
\limits_{\rho=1}^{\tilde q}\tilde\epsilon_\rho \frac{\delta
H_\rho}{\delta\boldsymbol{u}} D^{-1}(H_\rho) +\displaystyle\frac12
\sum\limits_{\alpha=1}^q \varepsilon_\alpha
\left(\displaystyle\frac{\delta K_\alpha}{\delta\boldsymbol{u}}
D^{-1}(\psi_\alpha)
+\displaystyle\frac{\delta\psi_\alpha}{\delta\boldsymbol{u}}
D^{-1}(K_\alpha) \right) \end{equation}
satisfies
$J \tilde
P J=\boldsymbol{\gamma}\,'-(\boldsymbol{\gamma}\,')^\dagger$.
\end{theorem}

The proof is by straightforward computation. Note that imposing the
conditions (\ref{loca0}) is a~very weak restriction, as
(\ref{loca0}) can be shown to follow from weak nonlocality and
symplecticity of $J\tilde P J$ under certain minor technical
assumptions.

The conditions (\ref{loca0}) have a very simple meaning. The f\/irst
of these conditions ensures that $L_{\boldsymbol{Y}_\rho}(J)=0$,
i.e., $\boldsymbol{Y}_\rho$ are Hamiltonian with respect to $J$. The
second condition means that the action of the operator $N=J \tilde
P$ on $\delta\psi_\alpha/\delta\boldsymbol{u}$
yields a variational derivative of another Hamiltonian density
$K_\alpha$. Moreover, if the operator $N^\dagger=\tilde P
J$ is hereditary, the said second condition guarantees \cite{oevth}
that $N^k (\delta\psi_\alpha/\delta\boldsymbol{u})$ are
variational derivatives (of possibly nonlocal Hamiltonian densities)
for all $k=2,3,\dots$.
\looseness=-1

Combining Theorem~\ref{hso} and Corollary~\ref{gam2} we readily
obtain the following results.

\begin{corollary}\label{gam3}Let $P$ be a nondegenerate Hamiltonian operator
such that $J=P^{-1}$ is weakly nonlocal and can be written in the
form \eqref{so} for suitable $p$, $q$, $b_m$ and $\psi_\alpha$. Then
under the assump\-tions of Theorem~{\rm \ref{hso}} any formally skew-symmetric
operator $\tilde P:\mathcal{V}^*\rightarrow\mathcal{V}$ such that $[P,\tilde P]{=}0$ can be written as
$\tilde P=L_{\boldsymbol{\tau}}(P)$, where
$\boldsymbol{\tau}=-P\boldsymbol{\gamma}$ and $\boldsymbol{\gamma}$
is given by \eqref{gameq2}.
\end{corollary}

\begin{corollary}\label{gam3loc}
Under the assumptions of Corollary~{\rm \ref{gam3}}
suppose that $P$ is a weakly nonlocal operator of the form \eqref{p}
and we have
\begin{equation}\label{loctau}
L_{\boldsymbol{G}_\rho}(\delta
K_{\alpha}/\delta\boldsymbol{u})=0,\qquad
L_{\boldsymbol{G}_\rho}(\delta
\psi_{\alpha}/\delta\boldsymbol{u})=0, \qquad \alpha=1,\dots,q,\quad
\rho=1,\dots,\bar q. \end{equation}
Then $\boldsymbol{\tau}=-P\boldsymbol{\gamma}$ is weakly nonlocal.

Moreover, if $P$ is a
differential operator then
$\boldsymbol{\tau}=-P\boldsymbol{\gamma}$ has the form
\begin{equation}\label{tau1} \boldsymbol{\tau}=\boldsymbol{\tau}_0
+\displaystyle\frac12 \sum \limits_{\rho=1}^{\tilde
q}\tilde\epsilon_\rho P\left(\displaystyle\frac{\delta
H_\rho}{\delta\boldsymbol{u}}\right) D^{-1}(H_\rho)
-\displaystyle\frac12 \sum\limits_{\alpha=1}^q \varepsilon_\alpha
P\left(\displaystyle\frac{\delta
K_\alpha}{\delta\boldsymbol{u}}\right) D^{-1}(\psi_\alpha),
\end{equation}
where $\boldsymbol{\tau}_0\in\mathcal{V}$ is local.
\end{corollary}
\begin{proof}
Using (\ref{loctau}) and Corollary~\ref{gam2} we readily see that
under the assumptions made
$\boldsymbol{\tau}=-P\boldsymbol{\gamma}$ is indeed weakly nonlocal.
If $P$ is a dif\/ferential operator then we have
$P\left(\displaystyle\frac{\delta\psi_\alpha}{\delta\boldsymbol{u}}\right)=0$
by Lemma~\ref{cas},
and a straightforward computation yields (\ref{tau1}).
\end{proof}

For instance, let $n=2$, and $\boldsymbol{u}=(u,v)^T$. Consider
\[
J=\left(\begin{array}{cc}0 & 1\\-1 &
0\end{array}\right)\qquad\mbox{and}\qquad \tilde
P=\left(\begin{array}{cc}D+ 2 v D^{-1}\circ v & -2 v D^{-1}\circ
u\\-2 u D^{-1}\circ v & D+2 u D^{-1}\circ u\end{array}\right),
\]
the symplectic structure and the Hamiltonian structure for the nonlinear
Schr\"odinger equation, see e.g.\ \cite{wang} and references
therein. We can rewrite $\tilde P$ as
\[\tilde P=
\left(\begin{array}{cc}D & 0\\0 &
D\end{array}\right)+\boldsymbol{Y}_1\otimes
D^{-1}\circ\boldsymbol{Y}_1,\qquad
\boldsymbol{Y}_1=\sqrt{2}\left(\begin{array}{c}-v\\
u\end{array}\right).
\]
We have
\begin{gather*}
J\tilde P J=\left(\begin{array}{cc}-D- 2 u D^{-1}\circ u & -2 u D^{-1}\circ v\\
-2 v D^{-1}\circ u & -D-2 v D^{-1}\circ
v\end{array}\right)=\left(\begin{array}{cc}-D & 0\\0 &
-D\end{array}\right) -\displaystyle\frac{\delta
H_1}{\delta\boldsymbol{u}}\otimes
D^{-1}\circ\displaystyle\frac{\delta H_1}{\delta\boldsymbol{u}},\\
H_1=(u^2+v^2)/\sqrt{2}.
\end{gather*}
The conditions of Theorem~\ref{gam0} and Corollary~\ref{gam3} are
readily seen to hold, and therefore we have
$J\tilde P J=\boldsymbol{\gamma}\,'-(\boldsymbol{\gamma}\,')^\dagger$, where
\[
\boldsymbol{\gamma}=\boldsymbol{\gamma}_0-\frac12
\displaystyle\frac{\delta H_1}{\delta\boldsymbol{u}}
D^{-1}(H_1),\qquad\boldsymbol{\gamma}_0=(v_1/2,u_1/2),
\]
and $\tilde P=L_{\boldsymbol{\tau}}(J^{-1})$, where
\[
\boldsymbol{\tau}=-\boldsymbol{u}_1/2+\frac12 \boldsymbol{Y}_1
D^{-1}(H_1).
\]

Given a Hamiltonian operator $P$, it
is natural to ask under which conditions $\tilde
P=L_{\boldsymbol{\tau}} (P)$ also is a Hamiltonian operator. A
straightforward but tedious computation yields the following
\begin{theorem}\label{hso2}Under the assumptions of Corollary~{\rm \ref{gam3}}
suppose that there exist local functions $L_\rho$ and $M_\alpha$
such that 
\begin{gather}\label{loca02} J \tilde P (\delta
H_\rho/\delta\boldsymbol{u}) =\delta
L_\rho/\delta\boldsymbol{u},\qquad\rho=1,\dots,\tilde q,\qquad\mbox{and} \\
 J \tilde P(\delta
K_\alpha/\delta\boldsymbol{u})=\delta M_\alpha/\delta\boldsymbol{u},
\qquad \alpha=1,\dots,q.\nonumber
\end{gather}

Then
$\tilde P=L_{\boldsymbol{\tau}} (P)$ is a Hamiltonian operator if
and only if
there exists a 
local $\widetilde{\boldsymbol{\gamma}}_0\in\mathcal{V}^*$ such that\looseness=-1
\begin{gather}
\widetilde{\boldsymbol{\gamma}}=\widetilde{\boldsymbol{\gamma}}_0
-\displaystyle\frac12 \sum \limits_{\rho=1}^{\tilde
q}\tilde\epsilon_\rho \left(\displaystyle\frac{\delta
L_\rho}{\delta\boldsymbol{u}} D^{-1}(H_\rho) +\frac{\delta
H_\rho}{\delta\boldsymbol{u}} D^{-1}(L_\rho)\right)\nonumber\\
\phantom{\widetilde{\boldsymbol{\gamma}}=}{}+\displaystyle\frac12\sum\limits_{\alpha=1}^q \varepsilon_\alpha
\left(\displaystyle\frac{\delta M_\alpha}{\delta\boldsymbol{u}}
D^{-1}(\psi_\alpha) +\displaystyle\frac{\delta K_\alpha}{\delta\boldsymbol{u}} D^{-1}(K_\alpha)
+\displaystyle\frac{\delta
\psi_\alpha}{\delta\boldsymbol{u}} D^{-1}(M_\alpha) \right),\label{gameq22}
\end{gather}
satisfies
$(J \tilde P)^2
J=\tilde{\boldsymbol{\gamma}}\,'-(\tilde{\boldsymbol{\gamma}}\,')^\dagger$.
\end{theorem}
\begin{proof} 
By Proposition~1 of \cite{serg1} the operator $\tilde
P=L_{\boldsymbol{\tau}}(P)$ is Hamiltonian if and only if
\begin{equation}\label{psc} [L_{\boldsymbol{\tau}}^2(P),P]=0. \end{equation} If $P$ is
nondegenerate then by Lemma~\ref{lemcomp} the condition (\ref{psc})
is equivalent to the requirement that the operator $J
L_{\boldsymbol{\tau}}^2(P) J$ be symplectic.
It is readily seen that
\[
J L_{\boldsymbol{\tau}}^2(P) J=J L_{\boldsymbol{\tau}}(\tilde P) J=
L_{\boldsymbol{\tau}}(J\tilde P J)+2 (J\tilde P)^2 J.
\]
In turn, as $J\tilde P J$ is symplectic, we have
$L_{\boldsymbol{\tau}}(J\tilde P J)=(J\tilde P
J\boldsymbol{\tau})'-(J\tilde P J\boldsymbol{\tau})'^\dagger$, and,
as
$\boldsymbol{\tau}=-P\boldsymbol{\gamma}=-J^{-1}\boldsymbol{\gamma}$,
where $\boldsymbol{\gamma}$ is given by (\ref{gameq2}), we obtain
\[
L_{\boldsymbol{\tau}}(J\tilde P J)=-(J\tilde P
\boldsymbol{\gamma})'+(J\tilde P \boldsymbol{\gamma})'^\dagger,
\]
so the operator $L_{\boldsymbol{\tau}}(J\tilde P J)$ is symplectic.

Hence the operator $J L_{\boldsymbol{\tau}}^2(P) J$ is symplectic if
and only if so is $(J\tilde P)^2 J$. By virtue of (\ref{loca02}) the
operator $(J\tilde P)^2 J$ is weakly nonlocal, so we can verify
its symplecticity using Theorem~\ref{gam0}, and the result follows.
\end{proof}

Combining Theorem~\ref{hso2} and Corollary~\ref{gam2} we obtain the
following
\begin{corollary}\label{gam3a}
Under the assumptions of Theorem~{\rm \ref{hso2}}
suppose that $P$ 
is normal, weakly nonlocal and
has the form \eqref{p}.
Further assume that
we have \begin{gather}
L_{\boldsymbol{G}_\rho}(\delta
H_{\sigma}/\delta\boldsymbol{u})=0,
\qquad L_{\boldsymbol{G}_\rho}(\delta
L_{\sigma}/\delta\boldsymbol{u})=0,
\qquad \rho=1,\dots,\bar q, \quad  \sigma=1,\dots,\tilde q,\nonumber\\
L_{\boldsymbol{G}_\rho}(\delta
K_{\beta}/\delta\boldsymbol{u})=0,
\qquad L_{\boldsymbol{G}_\rho}(\delta
M_{\beta}/\delta\boldsymbol{u})=0,\nonumber\\
\hspace*{40mm} L_{\boldsymbol{G}_\rho}(\delta
\psi_{\beta}/\delta\boldsymbol{u})=0,
\qquad \rho=1,\dots,\bar q, \quad  \beta=1,\dots,q,\label{loctau3}\\
L_{\boldsymbol{Y}_\rho}(\delta
H_{\sigma}/\delta\boldsymbol{u})=0,\qquad
L_{\boldsymbol{Y}_\rho}(\delta
K_{\alpha}/\delta\boldsymbol{u})=0,\nonumber\\
\hspace*{40mm}L_{\boldsymbol{Y}_\rho}(\delta
\psi_{\alpha}/\delta\boldsymbol{u})=0, \qquad \alpha=1,\dots,q,\quad
\rho,\sigma=1,\dots,\tilde q,\nonumber
\end{gather} Then $\tilde P$ is a Hamiltonian operator if and
only if there exists a
weakly nonlocal $\tilde{\boldsymbol{\tau}}\in\tilde{\mathcal{V}}$
such that
$L_{\boldsymbol{\tau}}^2(P)=L_{\tilde{\boldsymbol{\tau}}}(P)$.
\end{corollary}
\begin{proof} 
We readily f\/ind that
$\tilde{\boldsymbol{\tau}}=-P(-J\tilde P\boldsymbol{\gamma}+2
\tilde{\boldsymbol{\gamma}})+\boldsymbol{Q}=\tilde P\boldsymbol{\gamma}-2
P\tilde{\boldsymbol{\gamma}}+\boldsymbol{Q}$, where
$\boldsymbol{Q}\in\mathcal{V}$ because $P$ is normal. In complete analogy
with the proof of Corollary~\ref{gam2} we f\/ind that
the conditions (\ref{loctau3})
ensure that the coef\/f\/icients at the nonlocal variables in
$\tilde{\boldsymbol{\tau}}$ are local, and therefore
$\tilde{\boldsymbol{\tau}}$ is weakly nonlocal.
\looseness=-2
\end{proof}

\section{Local Hamiltonian operators of zero order}

Now assume that $J$ has the form \begin{equation}\label{lso} J=b_0,
\end{equation} where $b_0$ is an $n\times n$ matrix with entries from
$\mathcal{A}$.

A complete description of all symplectic operators of this form can
be found in \cite{msurv}. Namely, if $J$ (\ref{lso}) is
symplectic then we have \cite{msurv} \begin{equation}\label{lso1}
b_0=\sum\limits_{s=1}^n b_0^{(1,s)}(x,\boldsymbol{u})
u^s_1+b_0^{(0)}(x,\boldsymbol{u}), \end{equation} i.e., $b_0$
depends only on $x$, $\boldsymbol{u}$, $\boldsymbol{u}_1$ and, moreover,
is linear in $\boldsymbol{u}_1$. Of course, for $J$ (\ref{lso}) to
be symplectic the quantities $b_0^{(1,s)}$ and $b_0^{(0)}$ must
satisfy certain further conditions, see \cite{msurv} for details.
\looseness=-1

\begin{corollary}\label{gam6}Let $P$ be a nondegenerate
Hamiltonian operator
such that $J=P^{-1}$ has the form~\eqref{lso}.
Then 
any formally skew-symmetric differential operator $\tilde
P:\mathcal{V}^*\rightarrow\mathcal{V}$ such that $[P,\tilde P]=0$
can be written as $\tilde P=L_{\boldsymbol{\tau}}(P)$ for a local
$\boldsymbol{\tau}\in\mathcal{V}$. \looseness=-1
\end{corollary}
\begin{proof}
Indeed, by Corollary~\ref{gam3} we can take
$\boldsymbol{\tau}=-P\boldsymbol{\gamma}$ and $\boldsymbol{\gamma}$
given by (\ref{gameq2}) is now local.
\end{proof}
\begin{theorem}\label{gam7}Let $P$ be a nondegenerate Hamiltonian operator
such that $J=P^{-1}$ has the form~\eqref{lso}. Then a formally skew-symmetric
differential operator $\tilde P:\mathcal{V}^*\rightarrow\mathcal{V}$
is a Hamiltonian differential operator $\tilde
P:\mathcal{V}^*\rightarrow\mathcal{V}$ compatible with $P$ if and
only if there exist a local $\boldsymbol{\tau}\in\mathcal{V}$ and a
local $\tilde{\boldsymbol{\tau}}\in\mathcal{V}$ such that $\tilde
P=L_{\boldsymbol{\tau}}(P)$ and
$L_{\boldsymbol{\tau}}^2(P)=L_{\tilde{\boldsymbol{\tau}}}(P)$.
\end{theorem}
\begin{proof} The existence of a local
$\boldsymbol{\tau}\in\mathcal{V}$ such that $\tilde
P=L_{\boldsymbol{\tau}}(P)$ is immediate from Corollary~\ref{gam6}.

By Proposition~1 of \cite{serg1} the operator $\tilde
P=L_{\boldsymbol{\tau}}(P)$ is Hamiltonian if and only if
$[L_{\boldsymbol{\tau}}^2(P),P]=0$.
But by Corollary~\ref{gam6} the latter equality holds if and only if
there exists a local $\tilde{\boldsymbol{\tau}}\in\mathcal{V}$ such
that $L_{\boldsymbol{\tau}}^2(P)=L_{\tilde{\boldsymbol{\tau}}}(P)$,
and the result follows.
\end{proof}

\section[Local Hamiltonian operators of Dubrovin-Novikov type]{Local Hamiltonian operators of Dubrovin--Novikov type}

Assume now that $P$ is a Hamiltonian operator of Dubrovin--Novikov
type \cite{dn, dn84}, cf.\ also \cite{fer, fer2, fer3}, i.e., it is
a matrix dif\/ferential operator with the entries
\begin{equation}\label{dn0} {P}^{ij}=g^{ij}(\boldsymbol{u})D +
\sum_{k=1}^{n}b^{ij}_{k}(\boldsymbol{u})u_1^{k}, \end{equation} and
$\det g^{ij}\neq 0$, i.e., $P$, considered as formal series, is
nondegenerate.
\looseness=-1

An operator $P$ (\ref{dn0}) with $\det g^{ij}\neq 0$ is \cite{dn,
dn84} a Hamiltonian operator if and only if $g^{ij}$ is a
contravariant f\/lat (pseudo-\nobreak)Riemannian metric on an
$n$-dimensional manifold $M$ with local coordinates $u^i$ and
$b^{ij}_{k}= -\sum\nolimits_{m=1}^{n}g^{im}\Gamma^{j}_{mk}$, where
$\Gamma^{j}_{mk}$ is the Levi-Civita connection associated with
$g^{ij}$: $\Gamma^{k}_{ij}=(1/2)\sum_{s=1}^{n}g^{ks} (\partial
g_{sj}/\partial x^i+\partial g_{is}/\partial x^j-\partial
g_{ij}/\partial x^s)$. Here $g_{ij}$ is determined from the
conditions $\sum_{s=1}^{n}g^{ks}g_{sm}=\delta^k_m$, $k,m=1,\dots,n$.

Let us pass to the f\/lat coordinates 
$\psi^\alpha(\boldsymbol{u})$, $\alpha=1,\dots,n$, of $g_{ij}$. In
these coordinates $g^{ij}$ becomes a constant matrix $\eta^{ij}$,
where $\eta^{ij}=0$ for $i\neq j$ and $\eta^{ii}$ satisfy
$(\eta^{ii})^2=1$, $i,j=1,\dots,n$, and the Hamiltonian operator $P$
of Dubrovin--Novikov type associated with $g^{ij}$ takes the form
\begin{equation}\label{canon} P_{\rm can}^{ij}=\eta^{ij}D. \end{equation}

\begin{theorem}[\cite{magri_hd}]\label{cpbloc}
Let $P$ be a nondegenerate Hamiltonian operator of Dubrovin--Novikov type and
$\tilde P:\mathcal{V}^*\rightarrow\mathcal{V}$
be a purely differential formally skew-symmetric operator such
that \begin{equation}\label{loca2} \left\{\int \psi^\alpha dx,\int
\psi^\beta dx\right\}_{\tilde P}=0,\qquad \alpha,\beta=1,\dots,n,
\end{equation} where $\psi^\alpha=\psi^\alpha(\boldsymbol{u})$ are
flat coordinates for the metric $g_{ij}$ associated with $P$.
Then $[P,\tilde P]=0$ if and only if there exist
 a local $\boldsymbol{\tau}\in\mathcal{V}$ such that
$\tilde P=L_{\boldsymbol{\tau}}(P)$.
\end{theorem}

\begin{corollary}\label{ham}
Under the assumptions of Theorem~{\rm \ref{cpbloc}} suppose that
\begin{equation}\label{loca2a} \left\{\int \psi^\alpha dx,\int \psi^\beta
dx\right\}_{L_{\boldsymbol{\tau}}^2(P)}=0,\qquad
\alpha,\beta=1,\dots,n,
\end{equation}

Then $\tilde P$ is a Hamiltonian operator compatible with $P$ if and
only if there exists a local $\tilde{\boldsymbol{\tau}}
\in\mathcal{V}$ such that
$L_{\boldsymbol{\tau}}^2(P)=L_{\tilde{\boldsymbol{\tau}}}(P)$.
\end{corollary}
\begin{proof} If there exist local
$\boldsymbol{\tau},\tilde{\boldsymbol{\tau}} \in\mathcal{V}$ such
that $\tilde P=L_{\boldsymbol{\tau}}(P)$ and
$L_{\boldsymbol{\tau}}^2(P)=L_{\tilde{\boldsymbol{\tau}}}(P)$ then
by Proposition 3 of \cite{serg1} the operator $\tilde P$ indeed is a
Hamiltonian operator compatible with $P$.
On the other hand, 
the existence of $\boldsymbol{\tau}$ such that $\tilde
P=L_{\boldsymbol{\tau}}(P)$ is guaranteed by Theorem~\ref{cpbloc}.
Thus we only have to show that if the operator $\tilde P$ is
Hamiltonian then there exists a {\em local}
$\tilde{\boldsymbol{\tau}} \in\mathcal{V}$ such that
$L_{\boldsymbol{\tau}}^2(P)=L_{\tilde{\boldsymbol{\tau}}}(P)$.

By Proposition~1 of \cite{serg1} the operator $\tilde
P=L_{\boldsymbol{\tau}}(P)$ is Hamiltonian if and only if
$[L_{\boldsymbol{\tau}}^2(P),P]=0$. As (\ref{loca2a}) holds by assumption,
by Theorem~\ref{cpbloc}
we have $[L_{\boldsymbol{\tau}}^2(P),P]=0$
if and only if there exists a local
$\tilde{\boldsymbol{\tau}} \in\mathcal{V}$ such that
$L_{\boldsymbol{\tau}}^2(P)=L_{\tilde{\boldsymbol{\tau}}}(P)$, and
the result follows.
\end{proof}

For a simple example, let $n=1$, $\boldsymbol{u}\equiv u$, and let
$P=D$ and $\tilde P=D^3+2 u D+u_1$ be the f\/irst and the second Hamiltonian
structure of the KdV equation. We have \cite{magri_hd} $\tilde
P=L_{\boldsymbol{\tau}}(P)$ for $\boldsymbol{\tau}=-(u^2+u_2)/2$,
and it is readily seen that the conditions of Corollary~\ref{ham} are
satisf\/ied, so there exists a local $\tilde{\boldsymbol{\tau}}$ such
that $L_{\boldsymbol{\tau}}^2(P)=L_{\tilde{\boldsymbol{\tau}}}(P)$.
An easy computation shows that the latter equality holds e.g.\ for
$\tilde{\boldsymbol{\tau}}=-u_4/2-u_1^2/2+5 u^3/6$.

\section{Conclusions}

In the present paper we extended the homotopy formula (\ref{hom}) to
a large class of weakly nonlocal symplectic structures, see
Theorem~\ref{gam0} above. 
Besides the potential applications to the construction of nonlocal
extensions for the variational complex, this result enabled us to provide a
complete description for a large class of weakly nonlocal
Hamiltonian operators compatible with a given nondegenerate weakly
nonlocal Hamiltonian operator $P$ that possesses a weakly nonlocal
inverse (Corollaries~\ref{gam2}, \ref{gam3}, \ref{gam3loc}, and
\ref{gam3a}) or, more broadly, with a given weakly nonlocal
symplectic operator $J$ (Theorems~\ref{hso} and \ref{hso2}). These
results admit useful simplif\/ications for the case of zero- and
f\/irst-order dif\/ferential Hamiltonian operators, as presented in
Sections 4 and 5. In particular, in Section~5 we provide a simple
description for a very large class of local higher-order Hamiltonian
operators compatible with a given local Hamiltonian operator of
Dubrovin--Novikov type.
Note that f\/inding an ef\/f\/icient complete description of the nondegenerate weakly
nonlocal Hamiltonian operators with a weakly nonlocal inverse is an
interesting open problem,
because such operators would naturally generalize the Hamiltonian
operators (\ref{dn0}) of Dubrovin--Novikov type from Section~5 and
the zero-order local Hamiltonian operators from Section 4.
\looseness=-1

Thus, we extended 
the Lie derivative approach to the study of Hamiltonian operators
compatible with a given Hamiltonian operator $P$ from
f\/inite-dimensional Poisson structures \cite{serg1, smi} and
Hamiltonian operators of Dubrovin--Novikov type \cite{m1, serg1} to
the weakly nonlocal Hamiltonian ope\-rators of more general form.
An important advantage of this approach is that the 
vector f\/ields~$\boldsymbol{\tau}$ and~$\tilde{\boldsymbol{\tau}}$ in
general involve a considerably smaller number of unknown functions
than a generic formally skew-symmetric operator being a ``candidate" for
a Hamiltonian operator compatible with $P$, and the search for such vector f\/ields is often much
easier than
calculating directly the Schouten brackets involved,  cf.\ also the
discussion in \cite{serg1, smi}. This could be very helpful in
solving the classif\/ication problems like the following one: to
describe all weakly nonlocal Hamiltonian operators 
compatible with a given Hamiltonian operator $P$
and having a certain prescribed form.
\looseness=-1

\subsection*{Acknowledgements}

I am sincerely grateful to Prof.\ M.~B\l aszak and Drs.\ M.~Marvan,
E.V.~Ferapontov, M.V.~Pavlov and R.G.~Smirnov for stimulating
discussions. I am also pleased to thank the referees for useful
suggestions. \looseness=-2

This research was supported in part by the Czech Grant Agency (GA \v
CR) under grant No.~201/04/0538, by the Ministry of Education, Youth
and Sports of the Czech Republic (M\v SMT \v CR) under grant MSM 4781305904 and by
 Silesian University in Opava under grant IGS
1/2004. \looseness=-2

\pdfbookmark[1]{References}{ref}

\LastPageEnding


\begin{thebibliography}{99}

\footnotesize\itemsep=0pt

\bibitem{bl}
B\l aszak M.,
Multi-Hamiltonian theory of dynamical systems,
Springer, Heidelberg, 1998.



\bibitem{v2} Bocharov A.V. et al.,
Symmetries and conservation laws for dif\/ferential equations of
mathematical physics,  American Mathematical
Society, Providence, RI, 1999.

\bibitem{cooke} Cooke D.B.,
Compatibility conditions for Hamiltonian pairs,
{\it J. Math. Phys.} {\bf 32} (1991), no. 11,
3071--3076.

\bibitem{magri_hd}Degiovanni L., Magri F., Sciacca V.,
On deformation of Poisson manifolds of hydrodynamic type, {\it
Comm.\ Math.\ Phys.} {\bf 253} (2005), 1--24,
\href{http://arxiv.org/abs/nlin.SI/0103052}{nlin.SI/0103052}.

\bibitem{dor}Dorfman I.,
Dirac structures and integrability of nonlinear evolution
equations,
John Wiley \& Sons, Chichester, 1993.


\bibitem{dn}
Dubrovin B.A., Novikov S.P., Hamiltonian formalism of
one-dimensional systems of the hydrodynamic type and the
Bogolyubov--Whitham averaging method,
{\it Soviet Math. Dokl.} {\bf 27} (1983),
665--669.

\bibitem{dn84}
Dubrovin B.A., Novikov S.P., On Poisson brackets of hydrodynamic
type, {\it Soviet Math. Dokl.} {\bf 30} (1984), 651--654.

\bibitem{fer} Ferapontov E.V.,
Compatible Poisson brackets of hydrodynamic type, {\it J. Phys. A:
Math. Gen.} {\bf 34} (2001), 2377--2388,
\href{http://arxiv.org/abs/math.DG/0005221}{math.DG/0005221}.


\bibitem{fer2} Ferapontov E.V.,
Dif\/ferential geometry of nonlocal Hamiltonian operators of
hydrodynamic type,
{\it Funct. Anal. Appl.} {\bf 25} (1991), 195--204.

\bibitem{fer3} Ferapontov E.V.,
Nonlocal Hamiltonian operators of hydrodynamic type, dif\/ferential
geometry and applications, {\it Am. Math. Soc. Trans.} {\bf 170}
(1995), 33--58.


\bibitem{ff3} Finkel F., Fokas A.S.,
On the construction of evolution equations admitting a master
symmetry, {\it Phys. Lett.~A} {\bf 293} (2002),
36--44, \href{http://arxiv.org/abs/nlin.SI/0112002}{nlin.SI/0112002}.


\bibitem{ff}
Fuchssteiner B.,  Fokas A.S., Symplectic structures, their
B\"acklund transformations and hereditary symmetries, {\em Phys.~D} {\bf 4} (1981/82), no. 1, 47--66.

\bibitem{li}Lichnerowicz A.,
Les vari\'et\'es de Poisson et leurs alg\`ebres de Lie associ\'ees,
{\it J. Differential Geometry} {\bf 12} (1977), no. 2, 253--300.

\bibitem{magri}Magri F.,
A simple model of the integrable Hamiltonian equation, {\em J. Math.
Phys.} {\bf 19} (1978), 1156--1162.

\bibitem{mal}Maltsev A.Ya.,
Weakly nonlocal symplectic structures, Whitham
method and weakly nonlocal symplectic structures of
hydrodynamic type, {\it J. Phys. A: Math. Gen.} {\bf 38} (2005), 637--682,
\href{http://arxiv.org/abs/nlin.SI/0405060}{nlin.SI/0405060}.


\bibitem{mn}Maltsev A.Ya., Novikov S.P.,
On the local systems Hamiltonian in the weakly non-local Poisson
brackets, {\it Phys. D} {\bf 156} (2001), no.~1--2, 53--80,
\href{http://arxiv.org/abs/nlin.SI/0006030}{nlin.SI/0006030}.


\bibitem{s}
Mikhailov A.V., Shabat A.B., Sokolov V.V., The symmetry
approach to classif\/ication of integrable equa\-tions, in What is
Integrability?, Editor V.E.~Zakharov, Springer, New York, 1991,
115--184.


\bibitem{mik1}
Mikhailov A.V., Shabat A.B., Yamilov R.I.,
The symmetry approach to classif\/ication of nonlinear equations.
Complete lists of integrable systems, {\it Russ.\ Math.\ Surv.\/}
{\bf 42} (1987), no.~4, 1--63.

\bibitem{mik}
Mikhailov A.V., Yamilov R.I., Towards classif\/ication of
(2+1)-dimensional~integrable equations. Integrability conditions. I,
{\em J.~Phys.\ A: Math.\ Gen.} {\bf 31} (1998), 6707--6715.


\bibitem{msurv}Mokhov O.I., Symplectic and Poisson geometry on loop sapces
of manifolds in nonlinear equations, in  Topics in Topology and
Mathematical Physics, Editor S.P. Novikov, AMS, Providence, RI, 1995, 121--151,
\href{http://arxiv.org/abs/hep-th/9503076}{hep-th/9503076}.

\bibitem{m1}Mokhov O.I., Compatible
Dubrovin--Novikov Hamiltonian operators, Lie derivative and
integrable systems of hydrodynamic type, {\em Theoret. and Math.
Phys.} {\bf 133} (2002), no.~2, 1557--1564,
\href{http://arxiv.org/abs/math.DG/0201281}{math.DG/0201281}.

\bibitem{mo}Mokhov O.I.,
Compatible nonlocal Poisson brackets of hydrodynamic type and
related integrable hierarchies,
{\it Theoret. and Math. Phys.} {\bf 132} (2002), no.~1, 942--954,
\href{http://arxiv.org/abs/math.DG/0201242}{math.DG/0201242}.



\bibitem{oevth}
Oevel W., Rekursionmechanismen f\"ur Symmetrien und
Erhaltungss\"atze in Integrablen Systemen, Ph.D. Thesis, University
of Paderborn, Paderborn, 1984.

\bibitem{olv_eng2}
Olver P.J., Applications of Lie groups to dif\/ferential
equations, Springer, New York, 1993.

\bibitem{as-sb}
Sergyeyev A., On recursion operators and nonlocal symmetries of
evolution equations, in Proc. Sem. Dif\/f. Geom., Editor D.~Krupka,
Silesian University in Opava, Opava, 2000, 159--173,
\href{http://arxiv.org/abs/nlin.SI/0012011}{nlin.SI/0012011}.

\bibitem{serg1} Sergyeyev A., A simple way to make a
Hamiltonian system into bi-Hamiltonian one, {\it Acta Appl. Math.}
{\bf 83} (2004), 183--197, \href{http://arxiv.org/abs/nlin.SI/0310012}{nlin.SI/0310012}.

\bibitem{serg2} Sergyeyev A.,  Why nonlocal recursion operators produce local symmetries:
new results and applications, {\it J.~Phys.~A: Math.\ Gen.} {\bf
38} (2005), no.~15, 3397--3407, \href{http://arxiv.org/abs/nlin.SI/0410049}{nlin.SI/0410049}.

\bibitem{smi} Smirnov R.G.,
Bi-Hamiltonian formalism: a constructive approach, {\it Lett. Math.
Phys.} {\bf 41} (1997), 333--347.

\bibitem{sok}Sokolov V.V., On symmetries of evolution equations,
{\em Russ.\ Math.\ Surv.\/} {\bf 43} (1988), no.~5, 165--204.


\bibitem{vai}Vaisman I., Lectures on the geometry of Poisson manifolds,
Birkh\"auser, Basel, 1994.

\bibitem{wangth} Wang J.P., Symmetries and conservation laws of
evolution equations, Ph.D. Thesis,  Vrije
Universiteit van Amsterdam, Amsterdam, 1998.

\bibitem{wang} Wang J.P.,
A list of $1+1$ dimensional integrable equations and their
properties,
{\it J. Nonlinear Math. Phys.} {\bf 9} (2002), suppl.~1, 213--233.

\end{thebibliography}
\end{document}